# A high-performance reconstruction method for partially coherent ptychography

WENHUI XU,[1,2,†] SHOUCONG NING,[3,†] PENGJU SHENG,[1] HUIXIANG LIN,[1] ANGUS I KIRKLAND,[4,5] YONG PENG[6] AND FUCAI ZHANG[1,*]

[1]*Shenzhen Key Laboratory of Robotics Perception and Intelligence, and the Department of Electrical and Electronic Engineering, Southern University of Science and Technology, Shenzhen 518055, China*
[2]*Paul Scherrer Institut, 5232 Villigen PSI, Switzerland*
[3]*Department of Material Science and Engineering, National University of Singapore, Singapore 117583, Singapore*
[4]*The Rosalind Franklin Institute, Harwell Campus, Didcot OX11 0FA, UK*
[5]*Department of Materials, University of Oxford, Parks Road, Oxford OX1 3PH, UK*
[6]*School of Materials and Energy, Electron Microscopy Centre of Lanzhou University, Lanzhou University, Lanzhou 730000, China*
*[zhangfc@sustech.edu.cn](mailto:zhangfc@sustech.edu.cn)
†These authors contributed equally to this work

**Abstract:** Ptychography is now integrated as a tool in mainstream microscopy allowing quantitative and high-resolution imaging capabilities over a wide field of view. However, its ultimate performance is inevitably limited by the available coherent flux when implemented using electrons or laboratory X-ray sources. We present a universal reconstruction algorithm with high tolerance to low coherence for both far-field and near-field ptychography. The approach is practical for partial temporal and spatial coherence and requires no *prior* knowledge of the source properties. Our initial visible-light and electron data show that the method can dramatically improve the reconstruction quality and accelerate the convergence rate of the reconstruction. The approach also integrates well into existing ptychographic engines. It can also improve mixed-state and numerical monochromatisation methods, requiring a smaller number of coherent modes or lower dimensionality of Krylov subspace while providing more stable and faster convergence. We propose that this approach could have significant impact on ptychography of weakly scattering samples.

## 1. Introduction

Ptychography is an important extension of coherent diffraction imaging (CDI) that enables non-destructive imaging of complex valued extended objects [1,2]. The essence of diffraction is interference, and hence the coherence of the wavefield plays a key role in ptychography. If the extent of the illumination is more than twice the spatial coherence length or the detection integration time is longer than the temporal coherence time of the illumination, the visibility of interference features in the recorded diffraction patterns starts to degrade. For phase retrieval algorithms developed under the assumption of full temporal and spatial coherence of the illumination, a small deviation from this can cause a mismatch between the algorithmic estimate and experimental data, resulting in a degradation in reconstruction quality and convergence failure [3,4]. However, due to its data redundancy, ptychography has an improved tolerance to partial coherence than plane wave CDI [5]. Nonetheless, partial coherence is still a limiting factor when implementing ptychography with attosecond pulses [6,7], electrons [8,9], and laboratory X-ray [10] sources.

One way to circumvent this problem is to apply tight spectral and spatial filtering albeit at the cost of significant photon/electron flux loss. As a result, this approach requires a relatively long exposure time to collect sufficient photons for per resolution element, which increases the stability requirements of the experimental setup. For three-dimensional X-ray imaging of integrated circuits, it has been noted that the coherent X-ray flux poses a major limit on the imaging rate, the maximum measurable volume, and the achievable resolution [11,12]. Similarly, in scanning transmission electron microscopy (STEM), the study of two-dimension (2D) materials demands a low electron dose to prevent the onset of radiation damage. It has been experimentally demonstrated that the inevitable decoherence, even using state-of-the-art field-emission guns, prevents the realisation of sub-angstrom resolution using defocused probes under low-dose conditions [9].

In recent years, various numerical algorithms have been proposed to accommodate partial coherence issue in CDI. Most of these methods are based on Wolf's coherent-mode theory [13], where a partially coherent wave is modeled as a set of orthometric coherent modes that can propagate independently in free space. Hence, the recorded intensity is the weighted sum of the diffraction intensity of each mode [14]. For conventional CDI, the coherence properties of the source, i.e., the spatial coherence length and spectral distribution, need to be accurately characterised for high-quality reconstruction and the spatial and temporal decoherence can be resolved separately using a multi-modal algorithm [15], PolyCDI [16] or numerical monochromatisation [17].

In contrast, due to the data redundancy in ptychography, mixed-state reconstruction algorithms have been shown to enable simultaneous retrieval of the sample and the coherence property regardless of the experimental geometry and the underlying causes of any mixed states [18,19]. The "blurring" of diffraction patterns can also be modeled as a convolution of the coherent wave intensity with the Fourier spectrum of the complex coherence function of the illumination [20,21]. It has been demonstrated that the Fourier transform of the complex coherence function can be approximated as a simple Gaussian function with an adjustable standard deviation $\sigma$ [20,22], which can be used to compensate for the reduced fringe contrast caused by partial spatial coherence [23]. An algorithm incorporating the Richardson-Lucy deconvolution has also been shown to be capable of reconstructing both the object and coherence properties of the illumination wavefield from the partially coherent diffraction patterns [21].

In this work, we propose a universal reconstruction method that can compensate for partial temporal and spatial coherence without requiring a *priori* knowledge of the coherent properties of the source. This method is also applicable to both near-field and far-field regimes and can be easily integrated into the existing ptychographic retrieval engines, such as the extended ptychographic iterative engine (ePIE) [24], regularized PIE (rPIE) [25], difference map (DM) [26], or least-squares maximum-likelihood (LSQML) method [27]. When incorporated into the mixed-state and numerical monochromatisation methods, our method can accelerate the convergence rate and relax some restrictions. Significant gains in resolution and image quality are demonstrated using defocused four dimensional-STEM (4D-STEM) datasets and optical near-field and far-field datasets.

## 2. Theory and methods

In a typical ptychographic experiment, a specimen mounted on a translation stage is illuminated by a spatially confined probe formed from an aperture or a condenser lens system, and a detector located in the far field is used to record the diffracted intensities over all scan positions. In the case of electrons the stage translation is replaced by a shift on the illuminating probe. The recording of the detected diffraction patterns is subject to the decoherence arising from the broad spectrum and fluctuations of the probe, mixed states within the object of interest, and the finite point spread function of the detector. In addition, the distance between the object and detector has a substantial effect on the robustness to decoherence, as demonstrated previously [28]. For far-field ptychography using illumination formed from a spatially coherent monochromatic probe, the diffracted intensity at the $n^{\text{th}}$ ($n$=1, 2, …, $N$) scan position is defined as $I_n^c(\boldsymbol{k})$ where:

$$I_n^c(\boldsymbol{k}) = \left| \Im\left( O_n(\boldsymbol{r}) \cdot P_n(\boldsymbol{r}) \right) \right|^2, \tag{1}$$

in which $k$ and $r$ are reciprocal (detector) and sample space coordinates, $O_n(r)$ is the object transmission function, $P_n(r)$ is the probe function and $\mathfrak{J}$ denotes the Fourier propagator. When the probe is partially coherent, the diffraction pattern can be formed by convolving the coherent diffraction intensity with the spectrum of a normalised mutual coherence function (MCF), i.e., the Fourier transform of the complex degree of coherence [29]. In this case the partially coherent intensity $I_n^{pc}(k)$ is expressed as:

$$I_n^{pc}(k) = I_n^c(k) \otimes \hat{\gamma}(k), \qquad (2)$$

where $\hat{\gamma}(k)$ is the Fourier transform of the normalised MCF, which accounts for the effect of both temporal and spatial coherence.

The MCF is given by

$$\Gamma(r_1, r_2; \tau) = J(r_1, r_2) \gamma_\|(\tau), \qquad (3)$$

in which $\gamma_\|(\tau)$ is the complex degree of temporal (longitudinal) coherence and $J(r_1, r_2)$ is the mutual optical intensity (MOI).

Modelled by the generalized Schell model, the MOI becomes

$$J(r_1, r_2) = \varphi(r_1) \varphi^*(r_2) \gamma_\perp(r_1 - r_2), \qquad (4)$$

where $\varphi(r_1)$ and $\varphi(r_2)$ are simultaneously illuminating waves at two different spatial points in the sample plane, and $\gamma_\perp(r_1 - r_2)$ characterises the spatial (transverse) coherence. Substituting Eq. (4) into Eq. (3), the normalized MCF is given by

$$\overline{\Gamma}(r_1, r_2; \tau) = \gamma_\perp(r_1 - r_2) \gamma_\|(\tau). \qquad (5)$$

When the illuminating radiation is fully coherent, the normalized MCF is equal to unity. If the MCF can be measured in advance or retrieved during the phase retrieval process [21] for a partially coherent probe, $I_n^{pc}(k)$ can thus be modeled accurately. Here, we remove the requirement for an accurate MCF measurement by enforcing a simple Gaussian convolution operation to estimate the diffraction intensity and a complex constraint on the object function in the iteration process. As we will show, this method works well for both partially coherent far-field and near-field ptychography.

This low-coherence tolerant method can be integrated into the existing ptychographic engines. Here, we use a modified partially coherent ePIE (pc-ePIE) as an example as shown in Fig. 1. Compared to the traditional ePIE algorithm, the two key changes are highlighted in boxes 1 and 2 in Fig. 1. Based on guidance provided by K. A. Nugent, a two-dimensional (2D) Gaussian-Shell model can approximate an arbitrary complex coherence function effectively [3]. Hence, in our implementation, a normalized 2D Gaussian kernel $g$ is convolved with the amplitude $A_{\Psi_n(k)}$ of the diffracted wave estimate $\Psi_n(k)$ as:

$$A^G_{\Psi_n(k)} = A_{\Psi_n(k)} \otimes g, \qquad (6)$$

$$g = \exp\left[-\left(\frac{x^2}{2\sigma_1^2} + \frac{y^2}{2\sigma_2^2}\right)\right], \qquad (7)$$

where the constants $\sigma_1$ and $\sigma_2$ are blurring factors in the $x$ (horizontal) and $y$ (vertical) directions. The values of $\sigma_1$ and $\sigma_2$ (0.5-2.0 typically) depend on the radiation used, with a small value corresponding to a long coherence length.

As the sample and probe functions can be simultaneously retrieved in ptychography, the complementary information provided by the amplitude and phase of the sample transmittance can be used as an additional constraint. We have found that incorporating a complex constraint [30] can help prevent algorithmic stagnation and trapping in local minima especially in scenarios with severe loss of coherence, which is verified in Section 4.7.

We initially consider a sample with a complex transmission function given by:

$$T(\mathbf{r}) = A(\mathbf{r}) \exp[i\phi(\mathbf{r})], \tag{8}$$

where $A(\mathbf{r})$ and $\phi(\mathbf{r})$ are the scalar amplitude and phase of the sample, respectively.

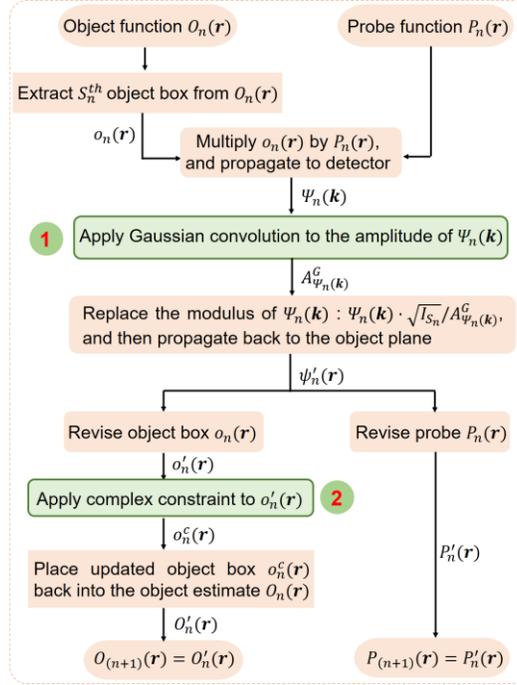

Fig. 1. Flowchart for one loop of the pc-ePIE algorithm. One entire iteration is composed of $N$ loops of this process. The recorded $N$ diffraction patterns are labeled as $I_{S_n}$ in a random sequence $S_n$. To accommodate all sample scan positions, the array size of $O_n(\mathbf{r})$ needs to be much larger than the array size of the extracted object box $o_n(\mathbf{r})$. The object box, probe, and diffraction patterns have the same array size.

In each loop of a pc-ePIE iteration, the phase $\phi'_n(\mathbf{r})$ and amplitude $A'_n(\mathbf{r})$ of the updated object function $o'_n(\mathbf{r})$ can be mutually constrained by (box 2 in Fig. 1):

$$A^c_n(\mathbf{r}) = \exp\left[(1-\alpha_1)\ln A'_n(\mathbf{r}) + \alpha_1 \overline{C_n} \phi'_n(\mathbf{r})\right], \tag{9}$$

$$\phi^c_n(\mathbf{r}) = (1-\alpha_2)\phi'_n(\mathbf{r}) + \alpha_2 \left(\overline{C_n}\right)^{-1} \ln A'_n(\mathbf{r}), \tag{10}$$

where the constants $\alpha_1$ and $\alpha_2 \in [0,1]$ are used to control the weighting of amplitude and phase and $\overline{C_n}$ is calculated over all $M$ pixels in the object box at every scan position:

$$\overline{C_n} = \frac{\sum_{m=1}^{M} |\ln A'_n(\mathbf{r})|}{\sum_{m=1}^{M} |\phi'_n(\mathbf{r})|}. \tag{11}$$

The resulting object transmission function after application of this complex constraint is $o^c_n(\mathbf{r}) = A^c_n(\mathbf{r}) \exp[i\phi^c_n(\mathbf{r})]$. The rationale and implementation details of the complex constraint are more thoroughly discussed in Section 4.8.

## 3. Experimental validation

### 3.1 Optical experiments

Optical experiments using a quantitative phase target (Benchmark QPT) were performed to validate the effectiveness of our proposed method. As shown in Fig. 2, a broadband and spatially incoherent white LED was used as light source, with relevant experimental parameters listed in Table 1. Bandpass spectral filters with different full widths at half maximum (FWHM) were used to control the temporal coherence, and aperture 1 with various diameters $D$ (30, 150, and 200 μm) acts as a spatial filter to adjust the spatial coherence length. For $D$=30 μm and FWHM=3 nm, reconstructed phases are shown in Figs. 3(b) and 4(b) and reach the theoretical resolution limit with high phase measurement precision. Consequently, we assume that the illumination is spatially and temporally quasi-coherent, and reconstructions from this setting were used as a reference for subsequent imaging quality comparisons. Aperture 2 was a 400-μm pinhole bonded to a piece of tissue, to create a structured wave that improves robustness to decoherence (see Section 4.3). The sample, QPT, mounted on a motorized $x/y$ stage, was scanned following a Fermat spiral trajectory [31]. The phase of QPT is related to the central wavelength $\lambda_c$ of the illumination, $\phi=2\pi/\lambda_c \cdot (n-1) \cdot d$, where $n$ (1.52 for QPT) is the refractive index and $d$ is the thickness of the object. Lens 4 located 35 mm upstream from the detector was used to achieve the Fraunhofer diffraction condition; in contrast, there was no optical element between the sample and detector in the near-field case. The diffraction patterns were recorded using a scientific camera with 6.5-μm pixels and a 16-bit dynamic range (Zyla 5.5 sCMOS, Andor). The total number of photons was kept approximately the same for all near-field or far-field experiments, which was achieved by adjusting the length of the exposure time.

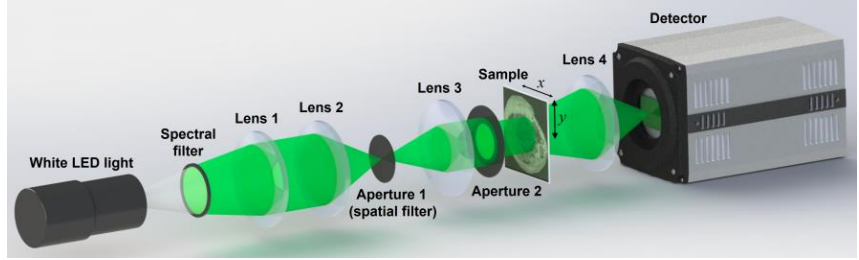

Fig. 2. Experimental geometry used for far-field partially coherent ptychography. Removing Lens 4 and placing the detector close to the sample, gives a setup for near-field ptychography.

**Table 1. Parameters of the QPT experimental configurations**

| Configurations | Near field | Far field |
| --- | --- | --- |
| FWHM of spectral filter | 3 nm/white light | 3/40/80 nm/white light |
| Central wavelength: $\lambda_c$ | 514.5/450 nm | 514.5/520/550/581.5 nm |
| Phase of the ROI | 1.30/1.49 rad | 1.30/1.29/1.22/1.15 rad |
| Diameter of aperture1: $D$ | 30/150/200 μm | 30/150 μm |
| Diameter of aperture2 | 400 μm | 400 μm |
| $f$ of lens1/ lens2/ lens3 | 75/50/45 mm | 75/50/45 mm |
| $f$ of lens4 | No | 35 mm |
| Exposure times | Single shot | Two shots |
| Aperture2-sample distance | 7.8 mm | 5.9/7.9 mm |
| Diffraction distance of sample exit-wave | 19.9 mm | ∞ |
| Size of diffraction patterns | 1024×1024 | 1024×1024 |
| Detector pixel size $\triangle k$ | 6.5 μm | 6.5 μm |
| Retrieved object pixel size $\triangle o$ | 6.5 μm | 2.71/2.73/2.89/3.06 μm |
| Scan step | 0.02 mm | 0.02 mm |
| Number of scan points | 100 | 100 |

Our proposed method is suitable for all ptychographic geometries with either low spatial or low temporal coherence. Firstly, near-field experiments with different temporal and spatial coherence were performed to validate the feasibility of pc-ePIE. There have been no solutions to partially temporal-coherent near-field ptychography except the mixed-state method yet. The specific broadband solutions (PloyCDI [16] and numerical monochromatisation [17]) can only be used for far-field CDI, since they utilise the field scaling of Fraunhofer diffraction. However, such a scaling does not exist for near-field diffraction. Here, we do not use the coherent-mode decomposition framework and solve the problem by introducing Gaussian convolution to the estimated diffraction modulus and complex constraints to the estimated object shown in Fig. 1.

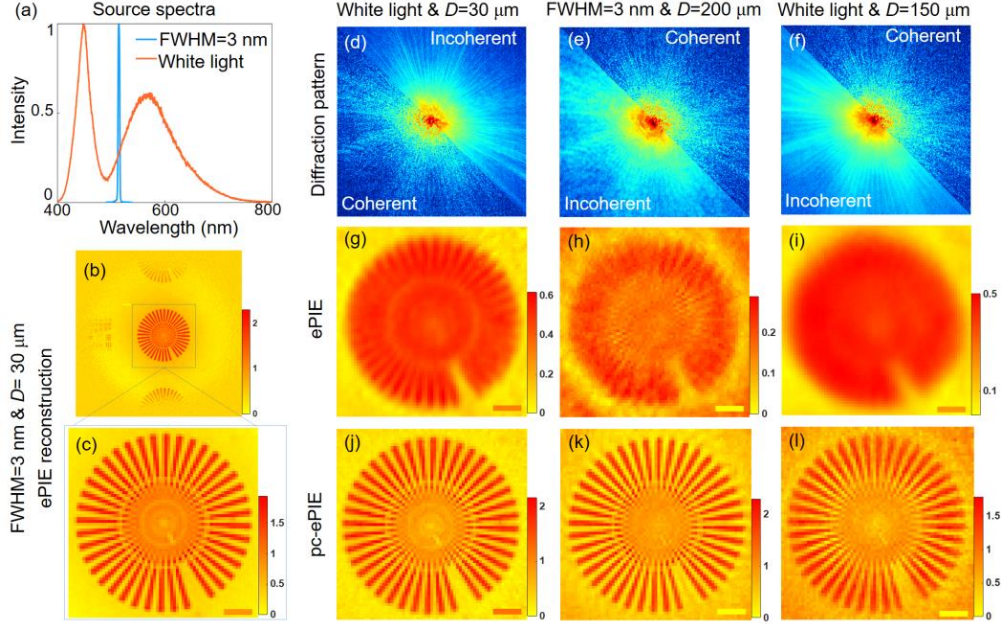

Fig. 3. Optical experimental validation in the near field. (a) Quasi-monochromatic spectrum (blue) and broadband spectrum (orange). (b) Reconstructed phase of the assumed coherent dataset from 100 iterations of ePIE. (c) Enlarged region marked in (b). (d-f) Diffraction patterns with different temporal and spatial coherence (on a logarithmic scale). Coherent parts are diffraction patterns with FWHM=3 nm and $D$=30 μm, and the incoherent parts are diffraction patterns with broadband (d), spatially decoherent (e), and simultaneously temporal and spatial decoherent (f) illumination. (g-i) Phase recovered by using ePIE algorithm. (j-l) Phase recovered using the pc-ePIE algorithm. (g) and (h) 500 iterations, (i) 1000 iterations, (j) and (k) 100 iterations and (l) 500 iterations. For pc-ePIE reconstructions, the values of $\sigma_1$ and $\sigma_2$ of Gaussian convolution were 1 for (j) and (k), and 1.5 for (l). The colorbar values are in rad for all phases displayed. Computation times for one iteration of ePIE and pc-ePIE were about 0.52 s and 0.64 s using an NVIDIA TITAN Xp. Scale bars are 65 μm.

To facilitate a description of quality degradation of the reconstructions from partially coherent datasets, we first performed experiments with a quasi-coherent probe generated using a spectral filter with FWHM=3 nm and a spatial filter (aperture 1) with $D$=30 μm. Figure 3(b) shows the retrieved sample phase using the ePIE algorithm. The enlarged image shown in Fig. 3(c) from the region marked in Fig. 3(b) indicates the region of interest (ROI) for subsequent comparison of reconstructed results. The ground-truth thickness difference $d$ over the ROI from QPT is 205 nm, corresponding to a phase of 1.30 radian at $\lambda_c$ =514.5 nm. This recovered phase under quasi-coherent illumination agrees with the ground-truth value.

For partially coherent illumination, the recorded diffraction patterns (labeled "Incoherent") exhibit obvious blur compared to the quasi-coherent patterns (labeled "Coherent"), as shown in Figs. 3(d)-3(f). The reconstructed phase using ePIE and pc-ePIE from the blurred diffraction patterns are shown in Figs. 3(g)-3(l). The pc-ePIE reconstructions show higher quality than ePIE, even for a fewer number of iterations. In the pc-ePIE reconstructions, the complex constraint was only used in the 5th-35th iterations for Figs. 3(j) and 3(k), and the 5th-35th and 105th-135th iterations for Fig. 3(l). The values of $\alpha_1$ and $\alpha_2$ were set to be 1 and 0 as QPT is a phase-dominant sample (as demonstrated in Section 4.7). This shows that a small number of iterations with complex constraints is sufficient to improve convergence speed. Alternatively, the Gaussian convolution was utilised in every iteration, with a Gaussian kernel array of 11×11 pixels and with $\sigma_1=\sigma_2=1$ for Figs. 3(j) and 3(k) and $\sigma_1=\sigma_2=1.5$ for Fig. 3(l). The selection of the array size and standard deviation of the Gaussian kernel is contingent upon the extent of blurring exhibited in the diffraction patterns due to decoherence set against the computational efficiency. When decoherence induces significant blurring in the diffraction patterns, larger array sizes and standard deviations are typically favored. However, increasing the kernel size escalates the computational load, potentially leading to impractical processing times. Therefore, we have carried out trial-and-error experiments to identify an effective size for the Gaussian kernel. In our reconstructions, we kept a constant Gaussian kernel size of 11×11 pixels, while adjusting $\sigma_1$ and $\sigma_2$ values to match variations in the degree of coherence. Compared to ePIE, the additional computation cost introduced by the complex constraint and Gaussian convolution of pc-ePIE is negligible.

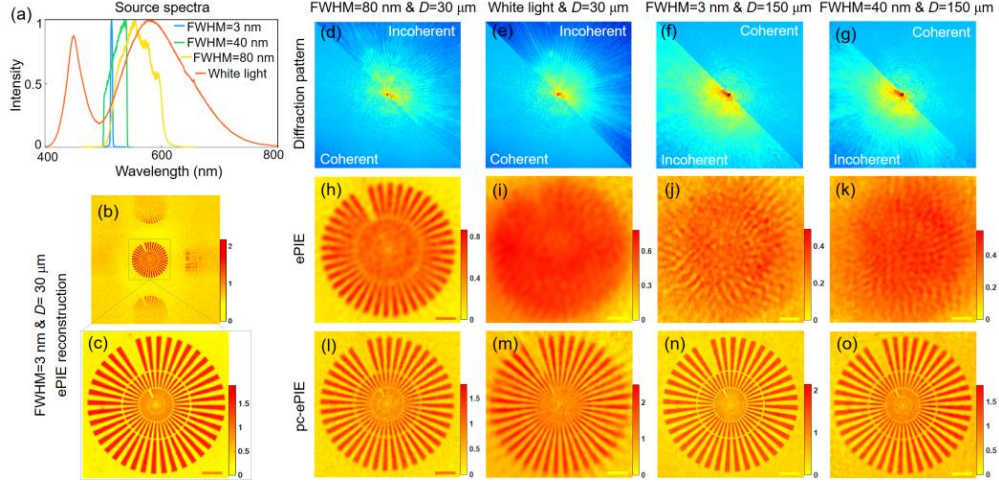

Fig. 4. Optical experimental validation in the far field. (a) Quasi-monochromatic spectrum (blue) and broadband spectra (green, yellow and orange). (b) Reconstructed phase from the assumed coherent dataset using 100 iterations of ePIE. (c) Enlarged region marked in (b). (d-g) Diffraction patterns (on a logarithmic scale) with different temporal and spatial coherence. The coherent parts are the diffraction patterns with FWHM=3 nm and D=30 μm, and the incoherent parts are the diffraction patterns with temporal (d) and (e), spatial (f), and the combination of temporal and spatial (g) incoherent illumination. (h-k) and (l-o) Phase recovered using ePIE and pc-ePIE, respectively. Reconstructed pixel sizes of (b), (h, l), (i, m), (j, n) and (k, o) were 2.71, 2.89, 3.06, 2.71 and 2.73 μm. The iterations for (h, i), (j, k), (l), (m), (n) and (o) were 1000, 1500, 100, 500, 350 and 400 respectively. The complex constraint was incorporated into pc-ePIE at the 5th-35th iterations for (l), 5th-35th, 105th-135th and 305th-335th iterations for (m), 5th-15th and 105th-115th iterations for (n), and 5th-35th and 105th-135th iterations for (o) with $\alpha_1=1$ and $\alpha_2=0$. Gaussian kernels used in pc-ePIE were all 11×11 pixels with $\sigma_1=\sigma_2=1$, convolved at every iteration. The computation times of one iteration of ePIE and pc-ePIE were about 450.7 ms and 452.2 ms using an NVIDIA TITAN Xp. All scale bars are 60 μm.

For the far-field case, to validate the versatility of our method, both coherent and partially coherent experiments were carried out using the spectra shown in Fig. 4(a). From Figs. 4(d)-4(e), the "scaling effect" related to the wavelength is clear in addition to the "blurring effect" caused by decoherence. When the spatial filters are of the same size $D$, the far-field distributions are the same except for the scaling related to wavelength. The scaling factor is given by $2\pi/(\lambda z)$, where $z$ is the Fraunhofer diffraction distance ($z$=35 mm, the same as the focal length of lens 4). However, phase retrieval approaches that utilise the "scaling effect" requires a precise measurement of the radiation spectrum in advance [16,17]. Without prior knowledge of the spectrum reconstruction of the temporally decoherent data using ePIE assuming full temporal coherence shows a severe reduction in quality for a FWHM=80 nm [Fig. 4(h)] and fails to converge when using the full spectrum of our lamp (denoted as white light) [Fig. 4(i)], in contrast to the quasi-coherent reconstruction shown in Fig. 4(c). Figures 4(l) and 4(m) show that the pc-ePIE method can enhance the retrieval quality even using fewer than half the iterations than used in the ePIE reconstructions. In addition, the pc-ePIE method can also overcome the limitations of spatial decoherence or a combination of spatial and temporal decoherence, as shown in Figs. 4(j), 4(k), 4(n) and 4(o).

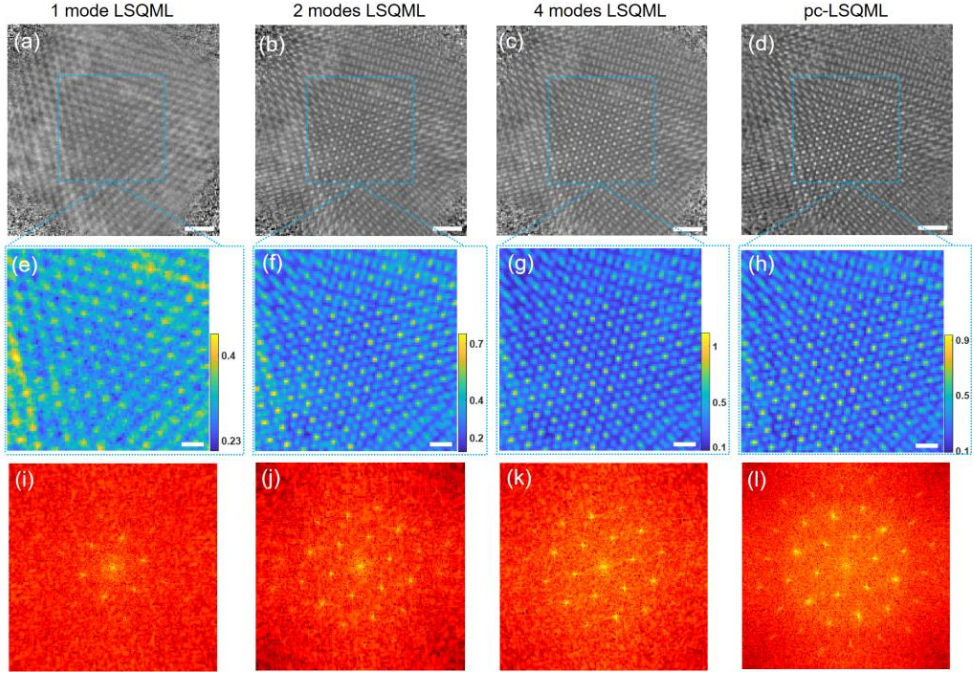

Fig. 5. Electron ptychographic reconstruction of a bilayer $MoSe_2/WS_2$ sample from 60×60 diffraction patterns. 1000 iterations were used for all reconstructions. Reconstructed real-space pixel size was 0.4 Å. (a-d) Reconstructed phase using different retrieval methods. Scale bars are 10 Å. (e-h) Enlarged images from the regions marked by the blue dashed rectangle in (a-d). Scale bars are 4 Å. (i-l) Corresponding diffractogram intensities (on a logarithmic scale) calculated from (a-d).

*3.2 Electron experiments*

To demonstrate the feasibility of our proposed method for electrons, we applied it to a partially coherent 4D-STEM dataset of a 2D multi-material sample, bilayer $MoSe_2/WS_2$, openly provided from Chen et al. [9]. The dataset was collected on an aberration-corrected Titan Themis instrument with a high-speed and high-dynamic range electron microscope pixel array detector (EMPAD) [32], using an 80keV beam energy, a 21.4-mrad probe-forming semi-angle, and ~50 nm probe defocus (see ref. [9] for more details). We have compared the reconstructions from single-mode (assuming full coherence) and mixed-state LSQML with our modified pc-LSQML for the same number of iterations, as shown in Fig. 5. For the pc-LSQML reconstruction, the kernel size was 11×11 pixels and the values of $\sigma_1$ and $\sigma_2$ were 0.8 for Gaussian convolution with the complex constraint incorporated at the $5^{th}$-$55^{th}$ iterations with $\alpha_1=1$ and $\alpha_2=0$. The determination criteria for $\alpha_1$ and $\alpha_2$ is detailed in Section 4.8. At least 4 probe modes are required for mixed-state LSQML to characterize the sample showing clear structural features of monolayer $WS_2$ and both well-aligned and misaligned stacking bilayer $MoSe_2/WS_2$ [Figs. 5(a)-5(c) and 5(e)-5(g)], and there was no further obvious improvement when additional modes are used. In contrast, the pc-LSQML can recover the sample directly without mode decomposition [Figs. 5(d) and 5(h)] and can retrieve higher spatial-frequency information than 4-mode LSQML [Figs. 5(k) and 5(l)]. In addition, the pc-LSQML method (~450 s) requires less time than the 4-mode LSQML (~530 s) for the 1000 iterations computed using an NVIDIA RTX A6000. Therefore, the reported method can efficiently compensate for partial coherence and reduce the computational requirements compared to the mixed-state method.

## 4. Discussion

### *4.1 Improvements over the mixed-state multi-slice ptychographic engine*

The method described can also optimise the mixed-state ptychographic engine. Here we take mixed-state multi-slice LSQML as an example and validate effectiveness by using a dataset of a $PrScO_3$ single crystal with a thickness of 21 nm projected along [001] provided by Chen et al. [33]. The dataset was also collected using an aberration-corrected electron microscope using an EMPAD detector, with a probe-forming aperture semi-angle of 21.4 mrad at 300 keV, a defocus of ~20 nm and scan step size of 0.41 Å (see ref. [33] for full experimental details). It has been shown that both a partially coherent treatment and multiple slices are required for high-quality reconstruction of a thick sample [33]. In our reconstructions, the sample was sliced into 21 slices, and the multi-slice retrieval method was integrated into both traditional LSQML and pc-LSQML engines. For the multi-slice pc-LSQML reconstruction, the settings of $\alpha_1$ and $\alpha_2$ were kept the same for all layers over all complex-constraint-incorporated iterations. The mean value of the calculated $\overline{C_n}$ over all layers and scan positions is close to 0.0. It is clear from Fig. 6 (a) that convergence errors reduce dramatically with an increase in the number of modes from 2 to 8, while they all rise after ~160 iterations for LSQML reconstructions. Our method leads to a stable reduction of error, and only two modes of pc-LSQML enable us to provide errors comparable to 8 modes of LSQML. Figures 6(b)-6(e) show that two probe modes are sufficient to resolve all heavy and light atoms for our pc-LSQML, whereas at least four modes are required for LSQML. Consequently, our method can reduce the required number of mixed states and also helps to stabilise convergence.

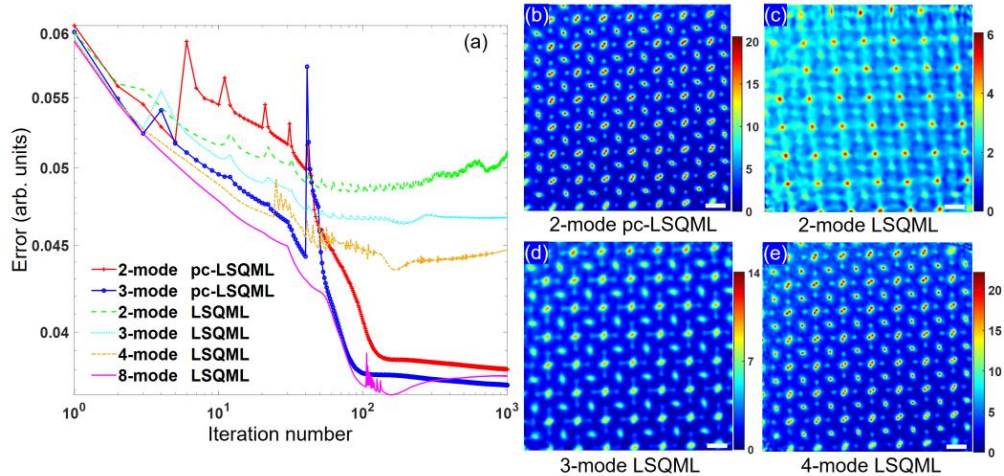

Fig. 6. Mixed-state multi-slice electron ptychographic reconstruction of $PrScO_3$ from 64×64 diffraction patterns. The reconstructed real-space pixel size was 0.0934 Å. (a) Convergence over 1000 iterations of pc-LSQML and LSQML algorithms with different probe eigenmodes. In the pc-LSQML reconstructions, $\sigma_1=\sigma_2=1$, $\alpha_1=1$ and $\alpha_2=0$, the complex constraint was incorporated at the 5th-45th iterations for the 2-mode case and at the 40th-80th iterations for the 3-mode case. Total phase summation of all slices from the reconstruction of (b) 2-mode pc-LSQML, (c) 2-mode LSQML, (d) 3-mode LSQML, and (e) 4-mode LSQML. Scale bars are 3 Å.

*4.2 Improvements over numerical monochromatisation*

Numerical monochromatisation [17] of polychromatic diffraction patterns is a very effective method for far-field broadband CDI. In this method, a monochromatic diffraction pattern is extracted from the broadband data by regularised inversion of a matrix $C$ that depends only on the radiation spectrum. The broadband diffraction **b** can be written as **b**=$C$**m**, where **m** represents the monochromatic pattern. Since the inversion is ill-conditioned (it is very sensitive to noise), a regularised method called conjugated gradient squares (CGLS) [34] can be used to mitigate this by minimizing the least-squares problem as:

$$\min_x \|Cx - b\|_2 \text{ subject to } x \in \kappa_k, \quad (12)$$

where $\kappa_k$ is the Krylov subspace:

$$\kappa_k \equiv \text{span}\{C^T b, C^T C C^T b, \cdots, (C^T C)^{k-1} C^T b\}. \quad (13)$$

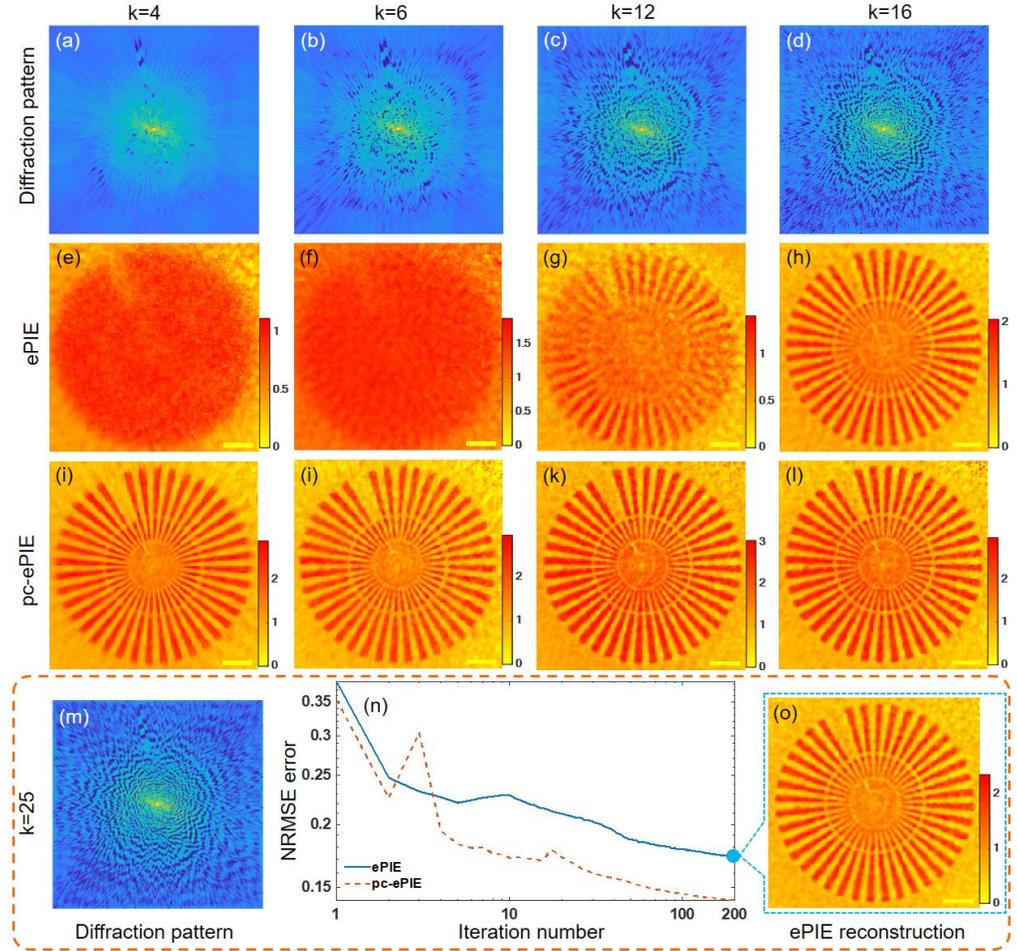

Fig. 7. Diffraction patterns (on a logarithmic scale) with various numerically monochromatised degrees and corresponding reconstructions using ePIE and pc-ePIE. The dataset was recorded with white light and $D$=30 μm. (a-d, m) Diffraction patterns monochromatised with $k$=4, 6, 12, 16, and 25. (e-h, o) and (i-l) Reconstructed phase using ePIE and pc-ePIE, respectively. (n) Convergence errors for ePIE and pc-ePIE over 200 iterations when $k$=25. The iteration numbers for (e-h) and (i-l) were 1000 and 200, respectively. In the pc-ePIE reconstructions, Gaussian convolution was utilised at every iteration with $\sigma_1=\sigma_2=1$; the complex constraint was incorporated at the 5$^{th}$-35$^{th}$ iterations with $\alpha_1=1$ and $\alpha_2=0$ for (i-l), and at the 2$^{nd}$-25$^{th}$ iterations with $\alpha_1=1$ and $\alpha_2=0.2$ for (o). Scale bars are 60 μm.

Typically, the monochromatic patterns are computed up to $k=40$, and the optimum value for $k$ is 25 chosen manually. Figure 7 shows the difference in diffraction patterns for different monochromatised degrees. The reconstruction quality using ePIE increases from $k=4$ to $k=25$ and reaches a maximum at $k=25$. For pc-ePIE the result with $k=6$ is comparable to the result recovered using ePIE at $k=25$ [Figs. 7(i) and 7(o)]. As a result, $k$ can be set to a value smaller than 25 in the monochromatisation process using the latter method, which represents a considerable time saving for a ptychographic dataset consisting of hundreds of diffraction patterns. Furthermore, the normalised root-mean-square-error (NRMSE) [35] calculated between the estimated and measured diffraction intensities of pc-ePIE is lower than that of ePIE, which indicates that pc-ePIE can model the incoherent diffraction wavefield more accurately [Fig. 7(n)].

*4.3 The influence of structured illumination on decoherence*

Structured illumination is used commonly to improve the reconstruction quality of CDI, since it can generate additional diversity in diffraction patterns compared to uniform illumination [36]. Here, we show that structured illumination can also compensate for decoherence. A large aperture can be segmented into several small regions by covering with a structured barrier. Consequently, the corresponding probe is split into sub-regions with a size closer to the coherence length than that of the uniform probe. It has been demonstrated that the influence of a broadband spectrum on ptychography can be mitigated by reducing the probe size [37], or by using structured illumination [38]. To extend this to both temporal and spatial decoherence, two far-field optical experiments (FWHM=40 nm and D=75 μm) were performed with probes produced using a tissue-covered pinhole [Fig. 8(a)] and a pure pinhole [Fig. 8(c)], respectively. From the phases recovered by ePIE shown in Figs. 8(b) and 8(d), it is clear that the reconstruction using structured illumination with 200 iterations shows more accurate structure than that from uniform illumination with 500 iterations.

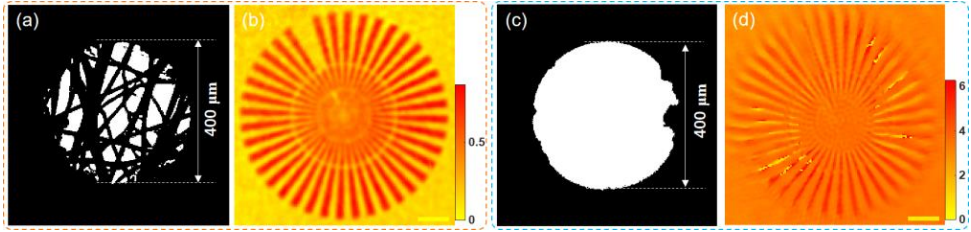

Fig. 8. Experimental aperture 2 and the corresponding reconstructed phase with FWHM=40 nm and $D$=75 μm. (a) Structured aperture 2 covered with a piece of tissue. (b) Phase retrieved for 200 iterations of ePIE. (c) Uniform aperture 2. (d) Phase retrieved for 500 iterations of ePIE. Scale bars in (b) and (d), 60 μm.

*4.4 Mixed-state reconstructions*

The mixed-state method [18] is the only current ptychographic solution to partially temporal and spatial coherence in both far-field and near-field. Therefore, we have also examined mixed-state LSQML reconstructions for comparison, as shown in Figs. 9 and 10. The mixed-state method [18] is based on coherent-mode theory and can model the probe and object incoherence precisely. LSQML [27] is an effective improvement of the maximum-likelihood ptychographic engine [39], which comprehensively accounts for system noise, scan position errors, and illumination wavefront variations with faster convergence.

Figure 9 shows near-field imaging results corresponding to Fig. 3. The mixed-state reconstructions [Figs. 9(c) and 9(f)] show comparable quality to the pc-ePIE reconstructions in Figs. 3(j) and 3(k) when the datasets were recorded with white light and $D$=30 μm, and with FWHM=3 nm and $D$=200 μm. However, the reconstructions using white light and $D$=150 μm [Figs. 9(g)-9(i)] are qualitatively worse than the pc-ePIE reconstruction shown in Fig. 3(l). We have also reconstructed with 14 and 20 modes and found that these additional modes do not improve the quality. For the far-field case, the mixed-state reconstructions [Figs. 10(c), 10(f), 10(i) and 10(l)] show higher resolution than the pc-ePIE reconstructions [Figs. 4(l)-4(o)], and both methods for white light are inferior to numerical monochromatisation [Fig. 7(o)].

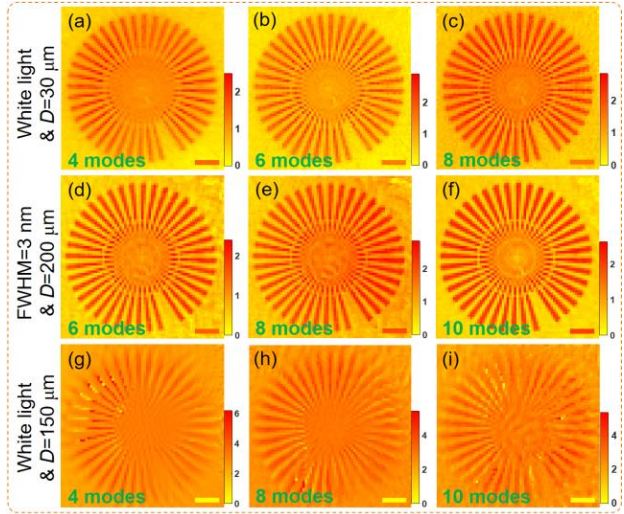

Fig. 9. Reconstructed phase from near-field datasets with (a-c) white light & $D$=30 μm, (d-f) FWHM=3 nm & $D$=200 μm, and (g-i) white light & $D$=150 μm using mixed-state LSQML for different probe modes indicated. Iterations for (a-f) and (g-i) are 1000 and 3000, respectively. Scale bars are 65 μm.

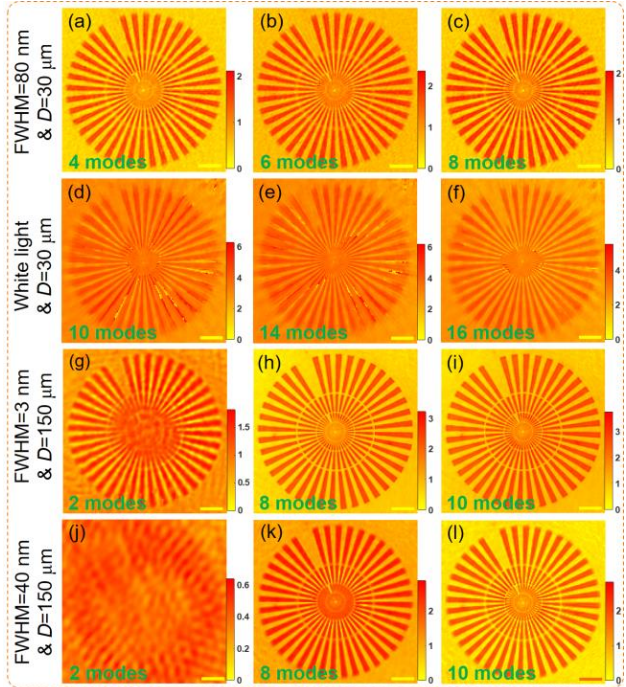

Fig. 10. Reconstructed phase from far-field datasets for (a-c) FWHM=80 nm & $D$=30 μm, (d-f) white light & $D$=30 μm, (g-i) FWHM=3 nm & $D$=150 μm, and (j-l) FWHM=40 nm & $D$=150 μm using mixed-state LSQML with different probe modes. Iterations for (a-c), (d-f), (g-j), and (k, l) are 1000, 3000, 2000 and 1500, respectively. Scale bars are 60 μm.

## 4.5 pc-rPIE and pc-DM reconstructions

We also validated that our method can be flexibly integrated into other ptychographic engines: rPIE [25] and DM [26], referred to subsequently as pc-rPIE and pc-DM. Figures. 11(a) and 11(b) show that traditional rPIE and DM methods cannot recover any phase profile even using 1000 iterations. In contrast, pc-rPIE and pc-DM give accurate reconstructions with no more than 400 iterations, as shown in Figs. 11(c) and 11(d). It is apparent that the phase recovered using pc-ePIE [Fig. 4(o)], pc-rPIE [Fig. 11(c)], and pc-DM [Fig. 11(c)] exhibit varying resolutions and require different iteration counts for convergence. These differences stem from the robustness and stability of ePIE, rPIE, and DM for different experimental setups and specimen types. The comparison between ePIE and rPIE has been previously reported [25], and our results here are consistent with these.

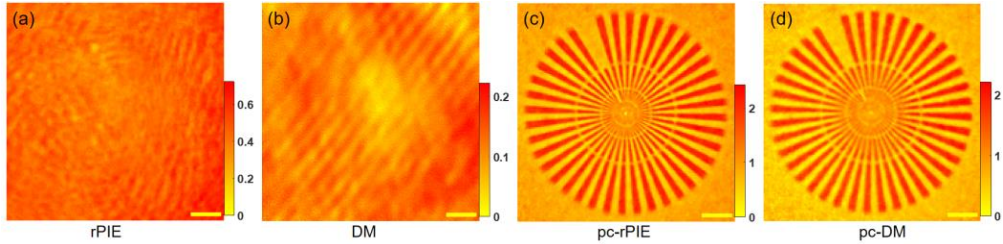

Fig. 11. Reconstructed phase from a far-field dataset with FWHM=40 nm and $D$=150 μm using different retrieval engines. Reconstructed pixel size was 2.73 μm. (a) and (b) Retrieved by rPIE and DM with 1000 iterations. (c) and (d) Retrieved by pc-rPIE and pc-DM with 250 and 400 iterations, respectively. A complex constraint was incorporated into pc-rPIE at $5^{th}$-$35^{th}$ and $205^{th}$-$235^{th}$ iterations (c), and pc-DM at $5^{th}$-$35^{th}$ iterations (d) with $α_1$=1 and $α_2$=0. Gaussian kernels used were all 11×11 pixels with $σ_1$=$σ_2$=1, being convolved at every iteration.

## 4.6 Amplitude-dominant object

As a representative amplitude object, a standard 2″ × 2″ positive 1951 USAF resolution target was used to test the efficacy of the proposed method. The datasets have high spatial coherence but low temporal coherence, selected from our previous work [28] which studied the influence of partial coherence on near- and far-field ptychography. See ref. [28] for full experimental details. In the reconstructions with pc-ePIE, the Gaussian convolution was utilised at every iteration with a kernel array of 11×11 pixels and $σ_1$=$σ_2$=1. The mean value of the calculated $\overline{C_n}$ over all scan positions is approximately 4.0, indicating that this USAF resolution target predominantly exhibits amplitude characteristics. Therefore, the complex constraint was performed with $α_1$=0 and $α_2$=1. Here, since the values of $α_1$ and $α_2$ for complex constraint were estimated roughly, this constraint was just applied at the $5^{th}$-$35^{th}$ iterations in every 100 iterations to accelerate the convergence. It is obvious from Fig. 12 that the spatial resolution can be improved significantly via pc-ePIE.

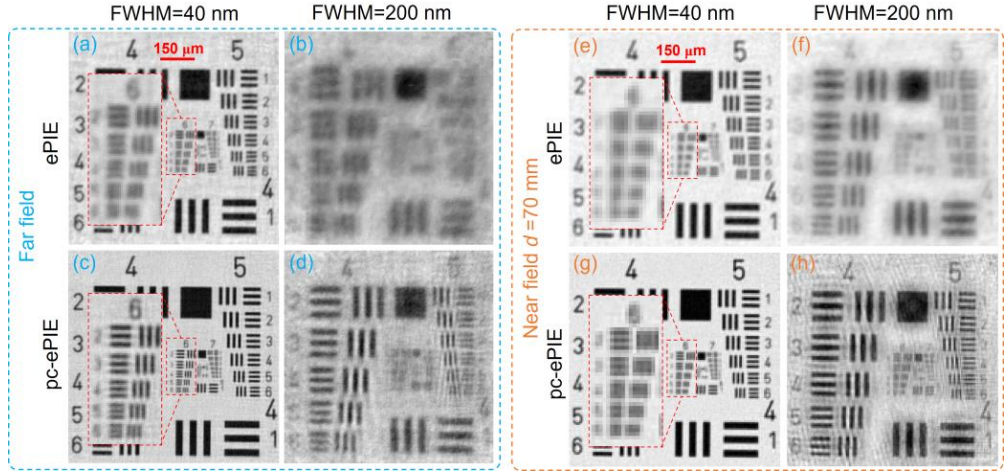

Fig. 12. Retrieved amplitudes under different spectral bandwidths and imaging configurations. The distance between the sample and detector is 70 mm for the near-filed imaging. Reconstructed pixel size: 6.5 μm for near field, 2.73 μm for FWHM=40 nm and 3.06 μm for FWHM=200 nm. (a), (b), (e) and (f) Retrieved using ePIE. (c), (d), (g) and (h) Retrieved using pc-ePIE. 100 iterations were used for (a), (c), (e) and (g), 1500 iterations for (b) and (d), 200 iterations for (f) and (h).

## 4.7 The Role of Complex constraint

The complex constraint serves to accelerate algorithm convergence, particularly in assisting escape from local minima and convergence to a global minima for conditions with severe degeneration of coherence, which has been validated using the results shown in Fig. 13. The Gaussian convolution here was utilised at every iteration, with a kernel array of 11×11 pixels and $\sigma_1=\sigma_2=1$. The complex constraint was used in the 5$^{th}$-35$^{th}$ iterations for Figs. 3(c) and 3(d) with $\alpha_1 = 1$ and $\alpha_2 = 0$. Compared to the results retrieved by complete pc-ePIE [Figs. 3(c) and 3(d)], removal of the complex constraint [Figs. 3(a) and 3(b)] significantly increases the number of iterations required for convergence. The mean value of $\overline{C_n}$ over all scan positions is 0.25, which shows that this QPT is a phase-dominant sample, consistent with the specification of QPT. This also shows that $\alpha_1$ and $\alpha_2$ can also be determined by $\overline{C_n}$ in visible-light regime.

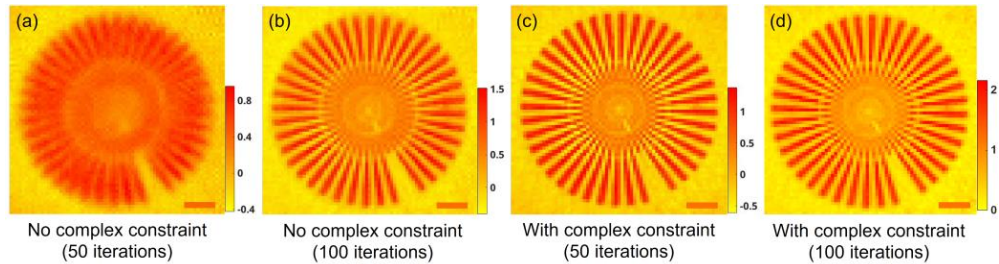

Fig. 13. Reconstructed phase using Gaussian convolution-enhanced ePIE. (a) and (b) Without complex constraint. (c) and (d) With complex constraint. The experimental data were collected in the near-field, under the same conditions as for Figs. 3(g) and 3(j). Scale bars are 65 μm.

*4.8 Setting the complex constraint*

The use of a complex constraint has been demonstrated to improve the quality of the resulting estimate of the specimen transmission function $T(\mathbf{r})$ by using complementary information provided by the phase $\phi(\mathbf{r})$ and magnitude $A(\mathbf{r})$ of the reconstructed wavefield in the X-ray regime [30]. In the projection approximation [40] for X-ray and extreme ultraviolet (EUV) regimes, $A(\mathbf{r})$ and $\phi(\mathbf{r})$ are related to the sample refractive index as:

$$n_s = 1 - \delta + i\beta, \tag{14}$$

$$A(\mathbf{r}) = \exp\left[-\frac{2\pi}{\lambda} \int_{sample} \beta(\mathbf{r},z) dz\right], \tag{15}$$

$$\phi(\mathbf{r}) = -\frac{2\pi}{\lambda} \int_{sample} \delta(\mathbf{r},z) dz, \tag{16}$$

where $z$ is the distance along the optical axis.

For a homogenous sample with a projected thickness $\tau(\mathbf{r})$,

$$A(\mathbf{r}) \rightarrow \exp[-2\pi/\lambda \cdot \beta \cdot \tau(\mathbf{r})], \tag{17}$$

$$\phi(\mathbf{r}) \rightarrow -2\pi/\lambda \cdot \delta \cdot \tau(\mathbf{r}). \tag{18}$$

Therefore, the amplitude and phase for homogenous samples or projected regions of the sample with the same material and density will have a fixed ratio $C(\mathbf{r})$ given by:

$$C(\mathbf{r}) = \frac{\ln A(\mathbf{r})}{\phi(\mathbf{r})} = \frac{\beta}{\delta}. \tag{19}$$

Theoretically, the values of $\alpha_1$ and $\alpha_2$ are directly linked through the ratio $\beta/\delta$; as for an increase in the ratio $\beta/\delta$, the value of $\alpha_1$ reduces and $\alpha_2$ increases. As knowledge of $\beta/\delta$ can be estimated from $\overline{C_n}$ [Eq. (11)] for the projected object in the iterative reconstruction process [30], $\alpha_1$ and $\alpha_2$ can also be determined from $\overline{C_n}$.

The relationship between the specimen transmission function and the complex refractive index, given by Eqs. (14)-(18), may not be universally applicable across other optical and electron wave regimes. However, using the complex constraint method, as in Eqs. (9)-(10), remains viable and applicable for ptychography. In the X-ray, EUV and visible-light regimes, the values of $\alpha_1$ and $\alpha_2$ can be easily determined by $\overline{C_n}$ for weakly scattering samples. Feasibility in the visible-light regime has been verified in Sections 3 and 4 and with a calculated value of $\overline{C_n}$, $\alpha_1$ and $\alpha_2$ can be set as described in Ref. [30].

For electrons, assuming that the specimen is weak-amplitude and weak-phase, $T(\mathbf{r})$ is defined as:

$$T(\mathbf{r}) = A(\mathbf{r})\exp[i\sigma_e v(\mathbf{r})], \tag{20}$$

where $v(\mathbf{r})$ is the projected atomic potential of the specimen, $\sigma_e$ is an interaction constant depending only on the energy of the incident electron beam, $A(\mathbf{r})$ arises from scattering outside of the detector or from electrons scattered by inelastic collisions [33,41]. For 2D materials, assuming the specimen is a weak phase object, $T(\mathbf{r})$ can be simplified as $T(\mathbf{r}) = \exp[i\sigma_e v(\mathbf{r})]$. This 2D object function represents the object transmissivity along the propagation direction. However, it requires that the specimen is thin enough to satisfy the assumption that the exit wavefield can be modelled as the product of a 2D object and the incident illumination. The thickness limit for this multiplication approximation is given as $t \leq 1.3\lambda/\theta_{max}^2$ [33,42], where $\lambda$ is the wavelength of the electron beam and $\theta_{max}$ is the maximum scattering angle recorded in a diffraction pattern. Beyond this limit, the resolution of the 2D ptychographic reconstruction will be affected by propagation and multiple scattering effects. In this case, the three-dimensional potential field can be divided into multiple slices. The projected potential is $v_i(\mathbf{r})$

($i = 1, 2, …, n$, $n$ is the number of slices) and the total projected potential can be numerically calculated by summing contributions over all slices:

$$\phi_{total}(\mathbf{r}) = \sum_{i=1}^{n}\phi_i(\mathbf{r}) = \sigma_e \sum_{i=1}^{n} v_i(\mathbf{r}), \tag{20}$$

where $\phi_i(\mathbf{r})$ is the phase of the $i$-th slice. For our reconstructions of 2D materials each slice of the thick specimen was treated as a weak phase object, making $\alpha_1=1$ and $\alpha_2=0$. In the multi-slice reconstructions, the complex constraint was performed at every slice with the same setting for $\alpha_1$ and $\alpha_2$.

It is worth noting that calculation of $\overline{C_n}$ is not required for each region of the sample that has a fixed value of $\beta/\delta$. In our reconstructions for both phase- and amplitude-dominant samples, as well as for the multi-material sample (a 2D bilayer $WS_2/MoS_2$ and a thick $PrScO_3$), we calculated $\overline{C_n}$ over all $M$ pixels in the object box at every scan position given by Eq. (11) and estimated the values of $\alpha_1$ and $\alpha_2$ roughly. Our experimental results demonstrate that the overlap constraint of ptychography enables a relaxation of the requirement for precise determination of $\alpha_1$ and $\alpha_2$, and eliminates any region segmentation for multi-material samples. Given that approximate estimates of $\alpha_1$ and $\alpha_2$, along with simple calculation of $\overline{C_n}$, tend to compromise the quantitative accuracy of the retrieved phase, typically only a small number of iterations with complex constraints can be used in the retrieval process. The range of iterations for complex constraint is not set at the end of the reconstruction, but rather at the outset of the iteration process or practically at the beginning of typically a few hundred iterations. The optimal time to switch on the complex constraint is after several iterations of a Gaussian-convolution-incorporated ptychographic engine to provide a reliable value for $\overline{C_n}$. A few iterations including a complex constraint can improve the convergence of the algorithm. Finally, given that the amplitude can be scaled, we set a threshold for $\overline{C_n}$ to confine its values within a range in our implementation, which enhances the stability and robustness of the algorithm.

## 5. Conclusions

We have demonstrated successful reconstructions for partially coherent near-field and far-field ptychographic datasets without *a priori* knowledge of the coherence properties of the radiation, providing a new route demonstrated for partial temporal and spatial decoherence of both electromagnetic and matter waves. Our method uses a simple Gaussian convolution to blur the estimated diffraction intensity and a complex constraint to update the amplitude and phase of the sample, which can be integrated into all ptychographic engines, including ePIE, rPIE, DM, and LSQML. We have successfully tested our modified pc-ePIE, pc-rPIE, pc-DM, and pc-LSQML using a series of partially temporal- and spatial-coherent near-field and far-field datasets in the visible-light regime, as well as for multi-material 4D-STEM datasets. Compared to traditional ptychographic engines, the modified algorithms significantly improve the reconstruction quality and convergence rate. In addition, the reported method can act as an optimisation for numerical monochromatisation and mixed-state methods. Ultrabroadband experiments with white-light illumination indicate that the method can accelerate the monochromatisation of large ptychographic datasets. Our results from a 4D-STEM dataset of a multi-material thick sample (consisting of both heavy and light atoms, $PrScO_3$) validate that combining the reported method with the mixed-state methods can dramatically reduce the number of required probe eigenmodes. This work opens up a new way to compensate for partial coherence for weakly scattering samples and as such will assist the development of high-throughput ptychography with X-rays or electrons and ultrafast experiments with broadband attosecond pulsed sources.

**Funding.** National Natural Science Foundation of China (12074167, 11775105), Shenzhen Key Laboratory of Robotics Perception and Intelligence (ZDSYS20200810171800001), Shenzhen Science and Technology Program

(KQTD20170810110313773), Centers for Mechanical Engineering Research and Education at MIT and SUSTech (6941806).

**Acknowledgments.** We acknowledge Prof. David A. Muller (Cornell University) and Prof. Zhen Chen (CAS, China) for sharing the 4D-STEM datasets, Prof. Zhen Chen for helpful discussions on electron ptychography, and Prof. Manuel Guizar-Sicairos (Paul Scherrer Institut, École Polytechnique Fédérale de Lausanne) and Dr. Abraham Lewis Levitan (Paul Scherrer Institut) for helpful discussions on the revision of the manuscript. We also appreciate the training supported by Prof. Yong Peng (Lanzhou University) and the training instructions from Dr. Kaiqi Bi (Lanzhou University).

**Data availability.** Visible-light data underlying the results presented in this paper are not publicly available at this time but may be obtained from the authors upon reasonable request.